\documentclass[preprint,12pt]{elsarticle}

\usepackage{amssymb}

\journal{$^*$Corresponding author. Email: yangli@iie.ac.cn}

\begin{document}

\begin{frontmatter}



\title{Qubit-string-based bit commitment protocols}

\author{Li Yang$^*$}
\author{Chong Xiang}
\author{Bao Li}
\address{State Key Laboratory of Information Security, Institute of Information Engineering, Chinese Academy of Sciences, Beijing 100195, China}

\begin{abstract}
Several kinds of qubit-string-based(QS-based) bit commitment protocols are presented, and a definition of information-theoretic concealing is given. All the protocols presented here are proved to be secure under this definition. We suggest an attack algorithm to obtain the local unitary transformation introduced in no-go theorem, which is used to attack the binding condition, then study the security of our QS-based bit commitment protocols under this attack via introducing a new concept "physical security of protocol". At last we present a practical QS-based bit commitment scheme against channel loss and error.
\end{abstract}




\end{frontmatter}
\newtheorem{theorem}{Theorem}
\newtheorem{lemma}[theorem]{Lemma}
\newtheorem{conjecture}[theorem]{Conjecture}
\newtheorem{corollary}[theorem]{corollary}
\newtheorem{definition}{Definition}
\newtheorem{proposition}[theorem]{Proposition}
\newtheorem{remark}{Remark}
\newtheorem{protocol}{Protocol}


\section{Introduction}\label{sec1}
Research on quantum cryptography may be traced back to about 40 years ago. Soon after Wiesner's work published \cite{Wie83}, Bennett and Brassard proposed two quantum cryptographical protocols in their original paper \cite{Ben84}: quantum key distribution (QKD) and quantum coin tossing. Though QKD had been proved unconditionally secure \cite{Ben92,Eke92,Eke91,Lo95,Deu96,May96_QKD,May01} and applied in practice, the quantum bit commitment (QBC) developed from quantum coin tossing has been proved impossible \cite{May97,Lo97}. A generally accepted QBC scheme was presented by Brassard, Crepeau, Jozsa and Langlois in 1993 \cite{Bra93}, but its unconditional security was shown to be impossible in 1996 \cite{May96_QBC}. Later, the idea in \cite{May96_QBC} was developed by Mayers \cite{May97} and Lo-Chau \cite{Lo97} independently and resulted in no-go theorem of QBC. It is shown that any kind of interactive protocol of QBC is also impossible \cite{Ari07}.

Although facing such clearly negative results, some authors still keep on exploring the unconditionally secure QBC which cannot be covered by the no-go theorem, or proving that the no-go theorem does not hold in some case. For example, Kent constructed a weaken scheme called quantum bit string commitment \cite{Ken03}, and then the concept of cheat-sensitive quantum bit commitment is presented by Hardy and Kent \cite{Har04}. These results were developed by Buhrman et al. \cite{Buh08}. Yuen believes that there generally exists unconditionally secure QBC protocols \cite{Yue03}, though his results have not been generally accepted yet.

In this paper, we show that the security of qubit-string-based (QS-based) bit commitment using the length of string as safe parameter is better than classical bit commitment, and it is possible to use QS-based bit commitment as a practical scheme. The paper is organized as follows: in Sec.\ref{sec2} some preliminaries are given; in Sec.\ref{sec3} we give the concept of information-theoretically concealing for quantum bit commitment; in Sec.\ref{sec4},\ref{sec5},\ref{sec6} three kinds of QS-based bit commitment protocols are presented and proved to be information-theoretically concealing; in Sec.\ref{sec7}, four other kinds of QS-based protocols are discussed; finally in Sec.\ref{sec8}, we show how to construct practical QS-based protocols against channel loss and error.

\section{Preliminaries}\label{sec2}
We relate here the concepts of classical bit commitment and $n_0^{th}$-order correlation immune Boolean functions, and describe a concrete form of EPR-attack suggested by the no-go theorem \cite{Lo97}, assuming that readers are familiar with the concepts of Boolean function and the content of no-go theorem of QBC.

\subsection{Bit commitment}
A bit commitment protocol includes two phases. In the commit phase, Alice determines a bit ($b$=0 or 1) and sends to Bob a piece of evidence. Later in the open phase, Alice opens the value of $b$ and some information of the evidence, and Bob checks whether Alice lies or not. A secure bit commitment needs two properties: binding and concealing. Binding means Alice cannot unveil $1-b$ without being detected after giving the evidence; concealing means Bob cannot get the value of $b$ before Alice unveils it. It can be proven that no classical bit commitment can satisfy both statistically concealing and statistically binding simultaneously.

After quantum cryptography being put forward, people desire to realize unconditionally secure QBC with quantum physics. Unfortunately, the no-go theorem of QBC \cite{May97, Lo97} says there cannot be unconditionally secure QBC protocol, only unconditionally concealing or binding protocols can be constructed.

\subsection{$n_0^{th}$-Order Correlation Immune Boolean Functions}
\begin{definition}
Let random binary variables $x_1,x_2,\ldots,x_n$ be independent and uniformly distributed. Then a Boolean function $f(x_1,\ldots,x_n): GF^n(2)\rightarrow GF(2)$ is called $n_0^{th}$-order correlation immune Boolean function if for every subset $\{i_1,\ldots,i_{n_0}\}\subset \{1,2,\ldots,n\}$, random variable $z=f(x_1,\ldots,x_n)$ is statistically independent of variable $(x_{i_1},\ldots,x_{i_{n_0}})$.
\end{definition}

\subsection{EPR-attack given in no-go theorem}\label{subsec1}
At the commitment phase of a QBC scheme, the committer Alice chooses commitment value $b$ towards the receiver Bob. For a cheating Alice, she can do as follows\cite{Lo97}:

\begin{enumerate}
\item Alice prepares a state $|\hat{0}\rangle$ without committing any values and sends the register $B$ to Bob,
\begin{eqnarray}\label{eqhat0}
|\hat{0}\rangle=\sum_i \alpha_i |e_i\rangle_A\otimes|\psi_i^{(0)}\rangle_B,
\end{eqnarray}
where $\langle e_i|e_j\rangle=\delta_{ij}$, but the normalized states $|\psi_i^{(0)}\rangle_B$ conform a set of nonorthogonal states.

\item At the open phase, if Alice decides to commit 0, she makes a measurement on the register $A$ and gets the value of $i$, then sends $i$ to Bob, and declares 0 as her commitment value.

\item if Alice decides to commit 1, she makes a local unitary operation $U_A$ on the register $A$ which satisfies:
\begin{eqnarray}
\langle\hat{1}|(U_A\otimes I)|\hat{0}\rangle=F(\mathrm{Tr}_A|\hat{0}\rangle\langle\hat{0}|,\mathrm{Tr}_A|\hat{1}\rangle\langle\hat{1}|)=1-\delta,
\end{eqnarray}
where
\begin{eqnarray}\label{eqhat1}
|\hat{1}\rangle=\sum_i\beta_i|e_i\rangle_A\otimes|\psi_i^{(1)}\rangle_B.
\end{eqnarray}
\end{enumerate}

Because the state $(U_A\otimes I)|\hat{0}\rangle$ is almost the same as the state $|\hat{1}\rangle$,  she can do as if she has sent the state$|\hat{1}\rangle$: she makes a measurement on the register $A$ and gets the value of $i$, and then tells Bob that she has committed the value 1 and sends $i$ to Bob. It can be seen that this attack strategy will be successful with probability $1-\delta$ with a small $\delta$.

\section{Information-theoretic security}\label{sec3}
In classical cryptography, the information-theoretic security is suggested by O. Goldrich \cite{Gold04} as follows:

\begin{definition}
A private key encryption is information-theoretically indistinguishable if for every circuit family \{$C_n$\}, every positive polynomial $p(\cdot)$, all sufficiently large $n$'s, and every $x, y$ in plaintext space:
\begin{eqnarray}
\Big|\mathrm{Pr}[C_n(E_{G(1^n)}(x))=1]-\mathrm{Pr}[C_n(E_{G(1^n)}(y))=1]\Big|<\frac{1}{p(n)},
\end{eqnarray}
where $G$ is a key generation algorithm.
\end{definition}

We suggest here a definition of information-theoretically concealing for quantum bit commitment protocol as follows:

\begin{definition}\label{def2}
A quantum bit commitment protocol is information-theoretically concealing if for every quantum circuit family \{$C_n$\}, every positive polynomial $p(\cdot)$, all sufficiently large $n$'s, and every $x,y\in\{0,1\}$:
\begin{eqnarray}\label{ITS}
\Big|\mathrm{Pr}[C_n(E_{G(1^n)}(x))=1]-\mathrm{Pr}[C_n(E_{G(1^n)}(y))=1]\Big|<\frac{1}{p(n)},
\end{eqnarray}
where the encryption algorithm $E$ should be a quantum algorithm.
\end{definition}

According to this definition, we can get the following theorem of concealing condition:

\begin{theorem}\label{the1}Let the density operators of quantum state Bob receives be $\rho_0$ and $\rho_1$, a QBC protocol is said to be information-theoretically concealing if for every positive polynomial $p(\cdot)$ and every sufficiently large $n$,
\begin{eqnarray}
D(\rho_0,\rho_1)<\frac{1}{p(n)}.
\end{eqnarray}
\end{theorem}
{\bf Proof.} Define $S_0$ as a set containing all the states Bob could receive when Alice commits 0. For every quantum circuit family $\{C_n\}$,
\begin{eqnarray}
& &\mathrm{Pr}[C_n(E_{G(1^n)}(0))=1]\nonumber\\
&= &\sum_{\rho_0^i\in S_0}p_i\cdot\mathrm{Pr}[C_n(\rho_0^i\otimes \sigma)=1]\nonumber\\
&= &\mathrm{Pr}[C_n(\sum_{\rho_0^i\in S_0}p_i\rho_0^i\otimes \sigma)=1]\nonumber\\
&= &\mathrm{Pr}[C_n(\rho_0\otimes \sigma)=1],
\end{eqnarray}
where $\sigma$ is the density operator of service bits of $C_n$.

Similarly,
\begin{eqnarray}
\mathrm{Pr}[C_n(E_{G(1^n)}(1))=1]=\mathrm{Pr}[C_n(\rho_1\otimes \sigma)=1].
\end{eqnarray}

Any quantum circuit family ${C_n}$ built for distinguishing two density operators corresponds to a set of positive operator-values measure (POVM) \{$E_m$\}. Define $p_m=\mathrm{Tr}(C_n(\rho_0\otimes \sigma)E_m)$, $q_m=\mathrm{Tr}(C_n(\rho_1\otimes \sigma)E_m)$ the probabilities of measurement outcomes labeled by $m$. In this case, we have:
\begin{eqnarray}
& &\Big|\mathrm{Pr}[C_n(\rho_0\otimes \sigma)=1]-\mathrm{Pr}[C_n(\rho_1\otimes \sigma)=1]\Big|\nonumber\\
&\leq & \max_{\{E_m\}}\frac{1}{2}\sum_m|\mathrm{Tr}[E_m(C_n(\rho_0\otimes \sigma)-C_n(\rho_1\otimes \sigma))]\nonumber\\
&= & \max_{\{E_m\}}D(p_m, q_m).
\end{eqnarray}
The last formula is equal to
\begin{eqnarray}
D(C_n(\rho_0\otimes \sigma),C_n(\rho_1\otimes \sigma))\leq D(\rho_0\otimes \sigma,\rho_1\otimes \sigma)=D(\rho_0, \rho_1)<\frac{1}{p(n)} . \end{eqnarray}
Hence, according to the Definition \ref{def2}, the theorem follows.$\hfill\Box$

To those QS-based protocols described in this paper, the safe parameter $n$ is the length of qubit string used in protocols.

\section{QS-based bit commitment based on coding of two non-orthogonal states\cite{Yan06}}\label{sec4}
\subsection{The scheme}
Let $|\psi_0\rangle$ and $|\psi_1\rangle$ be two non-orthogonal states, F($\cdot$) is an $n_0^{th}$-order correlation immune Boolean function. The protocol is as follows:
\begin{protocol}\label{pro1}~
\begin{enumerate}
\item Alice makes a commitment $b\in \{0,1\}$.
\item Alice chooses $a^{(i)}\in\{0,1\}^{n}$ randomly, here $i=1,2,\ldots,m$,  $a^{(i)}=(a_1^{(i)},a_2^{(i)},\ldots,a_n^{(i)})$ satisfies $F(a^{(i)})=b$.
\item Alice prepares $m\times n$ qubits in state $|\psi_{a_1^{(1)}}\rangle\cdots|\psi_{a_n^{(1)}}\rangle$ $|\psi_{a_1^{(2)}} \rangle\cdots|\psi_{a_n^{(2)}}\rangle\cdots\cdots\\|\psi_{a_1^{(m)}}\rangle \cdots |\psi_{a_n^{(m)}}\rangle$, and sends it to Bob as a piece of evidence for her commitment.
\item Alice opens by declaring {\it b} and the values of $a^{(i)}$.
\item Bob checks states of qubits by corresponding projective measurements: if $a_j^{(i)}=0$, Bob measures the $(n\times j-n+i)_{th}$ qubit with basis $\{|\psi_0\rangle, |\psi_0\rangle^{\bot}\}$; else with basis $\{|\psi_1\rangle, |\psi_1\rangle^{\bot}\}$. unless each results is matched, Bob has to break off the scheme.
\item Bob checks commitment value {\it b}. If $a^{(i)}$ satisfies $b=F(a^{(i)})$ for every $i$, Bob accepts the commitment value.$\hfill\Box$
\end{enumerate}
\end{protocol}

\subsection{The concealing condition}

 When $n_0=n-1$, $n_0^{th}$-order correlation immune Boolean function is the parity function
\begin{eqnarray}\label{TPF}
F(a^{(i)})=a_1^{(i)}\oplus a_2^{(i)}\oplus\cdots\oplus a_n^{(i)}.
\end{eqnarray}

Suppose density operator $\rho_b^{(n)}$ represents the state Bob receives when Alice commits $b$. As assumed, Alice sends each $|\psi_{a^{(i)}}\rangle$= $|\psi_{a_1^{(1)}}\rangle\cdots|\psi_{a_n^{(1)}}\rangle$ according to a uniform probability distribution, then
\begin{eqnarray}\label{eq3}
\rho_b^{(n)}= \frac{1}{2^{n-1}}\sum_{F(a^{(i)})=b}|\psi_{a^{(i)}}\rangle\langle \psi_{a^{(i)}}|.
\end{eqnarray}

\begin{lemma}\label{lem1}
The protocol \ref{pro1} is information-theoretically concealing.
\end{lemma}
{\bf Proof.} Let $\alpha$ be the angle between $|\psi_0\rangle$ and $|\psi_1\rangle$.

The quantum states $\rho_0^{(n)}$ and $\rho_1^{(n)}$ satisfy \cite{Ben96,Iva87}:
\begin{eqnarray}
\rho _0^{(n)}-\rho _1^{(n)}=2\times\left[
                            \begin{array}{cc}
                                    0 & \sin(\frac{\alpha}{2})\cos(\frac{\alpha}{2}) \\
                                    \sin(\frac{\alpha}{2})\cos(\frac{\alpha}{2}) & 0
                                    \end{array}
                                                                       \right]^{\otimes n} ,
\end{eqnarray}
then we have
\begin{eqnarray}\label{eq2}
D(\rho _0^{(n)},\rho _1^{(n)})=\frac{1}{2}\mathrm{Tr}\left|\rho _0^{(n)}-\rho _1^{(n)}\right|=(\sin\alpha)^n .
\end{eqnarray}

The parity function is usually used $m$ times in a scheme. We denote the density operator of these $m\times n$ qubits as $\rho_0^{(n,m)}$ and $\rho_1^{(n,m)}$. By using the triangle inequality of trace distance and $|A\otimes B|=|A|\otimes|B|$, we can show that
\begin{eqnarray}
D(\rho _0^{(n,m)}, \rho _1^{(n,m)})\leq m\times (\sin\alpha)^n.
\end{eqnarray}

It can be seen that for every given $m$, every positive polynomial $p(\cdot)$ and every sufficiently large $n$,
\begin{eqnarray}
D(\rho _0^{(n,m)},\rho _1^{(n,m)})<\frac{1}{p(n)} \label{eq0}
\end{eqnarray}
holds. According to Theorem 1, this lemma is proved. $\hfill\Box$

Our proof of inequality (\ref{eq0}) is valid only if parity function is used. We conjecture that if we use other $n_0^{th}$-order correlation immune Boolean functions to construct schemes, it may also satisfy inequality (\ref{eq0}).\\

\subsection{The Binding Condition}
The Mayers-Lo-Chau no-go theorem shows that while the bit commitment protocol is concealing, it can not be binding. Here we first show a concept of physical security of protocol, which means that the physical resource required in the breaking of a cryptosystem is beyond that of human beings given by the nature. Note that there is no protocol can achieve Shannon's computation security, the concept of physical security of protocol provides a way to reach Shannon's computation security.

In the \ref{appa}, we show a method to achieve the attack to the binding condition. Under such idea the attack algorithm's time complexity is $O(2^{3n})$, besides this algorithm needs at least $O(2^{2n})$ size of memory space to store the matrix. While $n=100$ the entry number of matrix $U_A$ is $2^{100}\times2^{100}$, this number is greater than the number of atoms of the earth(approximately $10^{50}$). It means that human beings cannot get the matrix actually, the attack strategy suggested in no-go theorem cannot be realized in this case forever, and our scheme may be physically secure on the binding side, if there is no efficient algorithm can help to find the local unitary transformation.

It has been proved that the security of classical bit commitment is at most statistically secure on one hand and computationally secure on the other hand, then the QS-based bit commitment with information-theoretically concealing and physically binding is a meaningful improvement if there is no efficient algorithm can help to find the local unitary transformation. However, whether the efficient algorithm exists is still an open problem.

Note that the parameter $m$ increases the trace distance between the density operators of the evidence for commit 0 and 1, it is used to resist another attack scheme toward binding condition. Every $a^{(i)}$ satisfies $F(a^{(i)})=0$ can become satisfying $F(a^{(i)})=1$ with one bit of change, and if Bob measures $|\psi_j\rangle$ with basis $\{|\psi_{(1-j)}\rangle,|\psi_{(1-j)}\rangle^{\bot}\}$, it takes probability $\frac{1}{2}$ that he accepts the result, therefore without $m$ Alice can cheat with a fifty-fifty chance of success, else she can cheat successfully only with a little probability $(\frac{1}{2})^m$. That is the reason why we add the parameter $m$.

\section{QS-based bit commitment based on conjugate coding\cite{Yan06}}\label{sec5}
\subsection{The scheme}
Let $|0\rangle_0=|0\rangle$, $|1\rangle_0=|1\rangle$, $|0\rangle_1=|+\rangle$, $|1\rangle_1=|-\rangle$, F($\cdot$) is an $n_0^{th}$-order correlation immune Boolean function. The protocol is as follows:
\begin{protocol}\label{pro2}~
\begin{enumerate}
\item Alice makes a commitment $b\in \{0,1\}$.
\item Alice chooses $a^{(i)}\in\{0,1\}^{n}$ randomly, here $i=1,2,\ldots,m$, $a^{(i)}=(a_1^{(i)},a_2^{(i)},\ldots,a_n^{(i)})$ satisfies $F(a^{(i)})=b$; and chooses $b^{(i)}\in\{0,1\}^{n}$ randomly, here $i=1,2,\ldots,m$, $b^{(i)}=(b_1^{(i)},b_2^{(i)},\ldots,b_n^{(i)})$.
\item Alice prepares $m\times n$ qubits in state $|a_1^{(1)}\rangle_{b_1^{(1)}}$ $\cdots|a_n^{(1)}\rangle_{b_n^{(1)}}|a_1^{(2)}\rangle_{b_1^{(2)}}$
    $\cdots|a_n^{(2)}\rangle_{b_n^{(2)}}$ $\cdots\cdots|a_1^{(m)}\rangle_{b_1^{(m)}}$ $\cdots |a_n^{(m)}\rangle_{b_n^{(m)}}$ and sends to Bob as a piece of evidence for her commitment.
\item Alice opens by declaring $b$ and the values of $a^{(i)}$ and $b^{(i)}$.
\item Bob checks states of qubits by corresponding projective measurements: if $b_j^{(i)}=0$, Bob measures with basis $\{|0\rangle, |1\rangle\}$, else with basis $\{|+\rangle, |-\rangle\}$.
\item Bob checks value $b$. If $a^{(i)}$ satisfies $b=F(a^{(i)})$ for every $i$, Bob accepts the value.$\hfill\Box$
\end{enumerate}
\end{protocol}

\subsection{The Concealing Condition}
Consider $F(\cdot$) is a parity function given in Eq.(\ref{TPF}). Define $\sigma_b^{(n)}$ the density operator of the state Bob receives when Alice commits $b$. Alice sends $|a^{(i)}\rangle_{b^{(i)}}=|a_1^{(i)}\rangle_{b_1^{(i)}}\cdots|a_n^{(i)}\rangle_{b_n^{(i)}}$, here $a^{(i)}$ satisfies $F(a^{(i)})=b$. For a uniform probability distribution we have
\begin{eqnarray}
\sigma_b^{(n)}=\frac{1}{2^{2n-1}}\sum_{b^{(i)}}\sum_{F(a^{(i)})=b}|a^{(i)}\rangle_{b^{(i)}}\langle a^{(i)}|.
\end{eqnarray}

Now we define two trace-preserving quantum operations $\mathcal{E}_1$ and $\mathcal{E}_2$.

Suppose $U_{\frac{\pi}{4}}^{\otimes n}$ is the operation element for $\mathcal{E}_1$, and $\{E_i\}$ is a set of operation elements for $\mathcal{E}_2$, here
\begin{eqnarray}
E_i=H^i=\frac{1}{2^n}H^{i_1}\otimes\cdots\otimes H^{i_n},
\end{eqnarray}
for $i\in\{0,1\}^n$. $U_{\frac{\pi}{4}}$ is a rotation operator, $H^0$ is the unit operator, and $H^1$ is the Hadamard operator.

Notice that
\begin{eqnarray}
H^jU_{\frac{\pi}{4}}^{\otimes n}|0\rangle_i=|i\rangle_{i\oplus\bar{j}},
\end{eqnarray}
here $i,j\in\{0,1\}^n$, and while $\alpha=\frac{\pi}{4}$ we have
\begin{eqnarray}
\rho_b^{(n)}=\frac{1}{2^{n-1}}\sum_{F(a^{(i)})=b}|0\rangle_{a^{(i)}}\langle0|.
\end{eqnarray}

Then we can get
\begin{eqnarray}
\mathcal{E}_2\circ\mathcal{E}_1(\rho_b^{(n)})&=&\mathcal{E}_2\bigg((U_{\frac{\pi}{4}})^{\otimes n}\rho_b^{(n)}((U_{\frac{\pi}{4}})^{\otimes n})^\dag\bigg) \nonumber\\
&=&\displaystyle{\sum_{j=1}^{2^n}}\sum_{F(a^{(i)})=b}\frac{1}{2^{2n-1}}H^j(U_{\frac{\pi}{4}})^{\otimes n}|0\rangle_{a^{(i)}}\langle0|((U_{\frac{\pi}{4}})^{\otimes n})^\dag (H^j)^\dag\nonumber\\
&=&\frac{1}{2^{2n-1}}\sum_{j=1}^{2^n}\sum_{F(a^{(i)})=b}|a^{(i)}\rangle_{a^{(i)}\oplus\bar{j}}\langle a^{(i)}|,
\end{eqnarray}
Let $b^{(i)}=a^{(i)}\oplus\bar{j}$, so
\begin{eqnarray}
\mathcal{E}_2\circ\mathcal{E}_1(\rho_b^{(n)})=\sigma_b^{(n)}.
\end{eqnarray}

Trace-preserving quantum operations are contractive, thus
\begin{eqnarray}
D(\sigma_0^{(n)},\sigma_1^{(n)})&=&D\Big(\mathcal{E}_2\circ\mathcal{E}_1(\rho_0^{(n)}),\ \mathcal{E}_2\circ\mathcal{E}_1(\rho_1^{(n)})\Big)\nonumber\\
&\leq& D(\rho _0^{(n)},\rho _1^{(n)}),
\end{eqnarray}
according to Eq. (\ref{eq2}), we have
\begin{eqnarray}
D(\sigma_0^{(n)},\sigma _1^{(n)})\leq(\sin\frac{\pi}{4})^n
\end{eqnarray}

\begin{lemma}  The protocol \ref{pro2} is information-theoretically concealing.
\end{lemma}
{\bf Proof.} As the $n$-variable parity function is reused $m$ times in our scheme, the two density operators of Bob's $m\times n$ qubits states are $\sigma_0^{(n,m)}$ and $\sigma_1^{(n,m)}$. By using the triangle inequality of trace distance and $|A\otimes B|=|A|\otimes|B|$, We can show that
\begin{eqnarray}
D(\sigma_0^{(n,m)},\sigma_1^{(n,m)})\leq m\times (\sin\frac{\pi}{4})^n.
\end{eqnarray}

It can be seen that for every given $m$, every positive polynomial $p(\cdot)$ and every sufficiently large $n$,
\begin{eqnarray}
D(\sigma _0^{(n,m)},\sigma _1^{(n,m)})<\frac{1}{p(n)} \label{eq1}
\end{eqnarray}
holds. Hence the lemma follows. $\hfill\Box$

We conjecture that if we use other $n_0^{th}$-order correlation immune Boolean functions instead of parity function to construct the scheme, it may satisfy the same inequality (\ref{eq1}).\\

\subsection{The Binding Condition}
It can be seen that the algorithm to solve $U_A$ in this case is also with $O(2^{3n})$ time complexity and at least $O(2^{2n})$ space complexity, then the binding condition of the protocol \ref{pro2} is the same as the protocol \ref{pro1}.

\section{QS-based bit commitment with referential bits}\label{sec6}
\subsection{The scheme}
\begin{protocol}\label{pro3}~
\begin{enumerate}
\item Alice makes a commitment $b\in \{0,1\}$.
\item Alice chooses $a^{(i)}$, $b^{(i)}$ and $c^{(i)}\in\{0,1\}^{n}$ randomly, here $i=1,2\cdots,m$, $a^{(i)}$ satisfies $F(a^{(i)})=b$.
\item Alice prepares $m\times 2n$ qubits in state $|a^{(1)}\rangle_{b^{(1)}}|c^{(1)}\rangle_{b^{(1)}}\cdots|a^{(m)}\rangle_{b^{(m)}}|c^{(m)}\rangle_{b^{(m)}}$, and sends to Bob with the values of $c^{(i)}$ published as a piece of evidence for her commitment.
\item Alice opens by declaring $b$ and the values of $a^{(i)}$ and $b^{(i)}$.
\item Bob checks states of qubits by corresponding projective measurements based on $b^{(i)}$ as the same as that of the protocol \ref{pro2}. Here he verifies two sets of data: first, the $c^{(i)}$ published before should accord with the measurement values; second, $a^{(i)}$ should satisfies $b=F(a^{(i)})$ for every $i$. If so, Bob accepts the commitment.$\hfill\Box$
\end{enumerate}
\end{protocol}

\subsection{The Concealing Condition}
Also consider $F(\cdot$) is a parity function given in Eq. (\ref{TPF}). For a uniform probability distribution, while the the $c^{(i)}$ is published, the density operator for every $i$ is
\begin{eqnarray}
\tau_b^{(n)}(c^{(i)})=\frac{1}{2^{2n-1}}\sum_{b^{(i)}}\sum_{F(a^{(i)})=b}
\left(|a^{(i)}\rangle_{b^{(i)}}\langle a^{(i)}|\otimes|c^{(i)}\rangle_{b^{(i)}}\langle c^{(i)}|\right),
\end{eqnarray}

then the trace distance between $\tau_0^{(n)}(c^{(i)})$ and $\tau_1^{(n)}(c^{(i)})$ is
\begin{eqnarray}
& &D(\tau_0^{(n)}(c^{(i)}),\tau_1^{(n)}(c^{(i)}))\nonumber\\
&=&\frac{1}{2^{2n}}\mathrm{Tr}\left|\bigotimes_{j=1}^{n}
\left(|0\rangle_0\langle0|\otimes|c_j^{(i)}\rangle_0\langle c_j^{(i)}|-
|1\rangle_0\langle1|\otimes|c_j^{(i)}\rangle_0\langle c_j^{(i)}|+\right.\right.\nonumber\\
& &~~~~~~~~~~~~~~\left.\left.
+|0\rangle_1\langle0|\otimes|c_j^{(i)}\rangle_1\langle c_j^{(i)}|
-|1\rangle_1\langle1|\otimes|c_j^{(i)}\rangle_1\langle c_j^{(i)}|\right)\right|.
\end{eqnarray}

Let
\begin{eqnarray}
\vartheta(i)=|0\rangle_0\langle0|\otimes|i\rangle_0\langle i|-
|1\rangle_0\langle1|\otimes|i\rangle_0\langle i|+\nonumber\\
+|0\rangle_1\langle0|\otimes|i\rangle_1\langle i|
-|1\rangle_1\langle1|\otimes|i\rangle_1\langle i|,
\end{eqnarray}

then we can rewrite the trace distance as
\begin{eqnarray}
D(\tau_0^{(n)}(c^{(i)}),\tau_1^{(n)}(c^{(i)}))=\frac{1}{2^n}\mathrm{Tr}\left|\bigotimes_{j=1}^{n}\vartheta(c_j^{(i)})\right|
=\frac{1}{2^{2n}}\prod_{j=1}^n\mathrm{Tr}\left|\vartheta(c_j^{(i)})\right|.
\end{eqnarray}

\begin{remark}
This direct product decomposition can be also used to solve the trace distance $D(\rho_0^{(n)},\rho_1^{(n)})$ and $D(\sigma_0^{(n)},\sigma_1^{(n)})$ of the first two protocols. We give that:
\begin{eqnarray}
\rho_0^{(n)}-\rho_1^{(n)}=\frac{1}{2^{n-1}}\left(|\psi_0\rangle\langle\psi_0|-|\psi_1\rangle\langle\psi_1|\right)^{\otimes n};
\end{eqnarray}
\begin{eqnarray}
\sigma_0^{(n)}-\sigma_1^{(n)}=\frac{1}{2^{2n-1}}\left(|0\rangle_0\langle0|-|1\rangle_0\langle1|+
|0\rangle_1\langle0|-|1\rangle_1\langle1|\right)^{\otimes n}.
\end{eqnarray}
In this way the trace distances can result in exact values.$\hfill\Box$
\end{remark}

The matrix expressions of $\vartheta(c_j^{(i)})$ are shown as:
\begin{eqnarray}
\vartheta(0)=\left[\begin{array}{rrrr}
               1 & 0 & 1/2 & 1/2 \\
               0 & 0 & 1/2 & 1/2 \\
               1/2 & 1/2 & -1 & 0 \\
               1/2 & 1/2 & 0 & 0
             \end{array}\right],\\
\vartheta(1)=\left[\begin{array}{rrrr}
               0 & 0 & 1/2 & -1/2 \\
               0 & 1 & -1/2 & 1/2 \\
               1/2 & -1/2 & 0 & 0 \\
               -1/2 & 1/2 & 0 & -1
             \end{array}\right].
\end{eqnarray}
They have the same eigenpolynomial as:
\begin{eqnarray}
\lambda^4-2\lambda^2+\frac{1}{4},
\end{eqnarray}

then we can have:
\begin{eqnarray}
\mathrm{Tr}\left|\vartheta(0)\right|=\mathrm{Tr}\left|\vartheta(1)\right|=2\sqrt{3},
\end{eqnarray}

so we get the value of the trace distance:
\begin{eqnarray}
D(\tau_0^{(n)}(c^{(i)}),\tau_1^{(n)}(c^{(i)}))=(\frac{\sqrt{3}}{2})^n,
\end{eqnarray}
it holds for every $i$ and $c^{(i)}$.

As the density operator for Bob while Alice commits $b$ is shown as:
\begin{eqnarray}
\tau_b^{(n,m)}=\bigotimes_{i=1}^m\tau_b^{(n)}(c^{(i)}).
\end{eqnarray}

The trace distance between $\tau_0^{(n,m)}$ and $\tau_1^{(n,m)}$ is easily given out:
\begin{eqnarray}
D(\tau_0^{(n,m)},\tau_1^{(n,m)})\leq m\times(\frac{\sqrt{3}}{2})^n,
\end{eqnarray}
which means
\begin{eqnarray}
D(\tau_0^{(n,m)},\tau_1^{(n,m)})\leq \frac{1}{p(n)}
\end{eqnarray}
can be held for every given m, every positive polynomial $p(\cdot)$ and every sufficiently large $n$. Based on the Theorem \ref{the1}, we know that this protocol is information-theoretically concealing.

\subsection{The Binding Condition}
It is the same as the above protocols.

\section{Other protocols}\label{sec7}
Besides three protocols described above, we can also construct other QS-based bit commitment protocols. There are four examples.

\subsection{Scheme using both variable states and function value states}
\begin{protocol}\label{pro4}~
\begin{enumerate}
\item Alice chooses a commitment value $b\in \{0,1\}$.
\item Alice chooses $x\in\{0,1\}^{n}$ randomly.
\item Alice prepares states $|0\rangle_x|0\rangle_y$, $y=f_b(x)$, then Alice sends the state to Bob as a piece of evidence for her commitment. Functions $f_0(\cdot), f_1(\cdot)$ are known by both of them.
\item Alice opens by declaring {\it b} and the values of $x$, Bob checks the received states of qubits.
\item Bob accepts the commitment if $y$ is equal to $f_b(x)$.$\hfill\Box$
\end{enumerate}
\end{protocol}

If we use only function value state $|0\rangle_y$ to commit, Alice can easily cheat via finding a collision. Then we use both variable states and function value states to commit. If Alice plans to cheat, she needs to prepare the state in Bob's hand remotely. According to the no-go theorem, she can prepare an entangled state
\begin{eqnarray}
\sum_x\left(|x\rangle|y\rangle\right)_A\otimes\left(|0\rangle_x|0\rangle_y\right)_B.
\end{eqnarray}

The concealing condition of this protocol is not easy to satisfy. We must guarantee that there is no simple correlation between variable bit and function value bit. It can be seen that permutation cannot be used in this protocol.

\subsection{Scheme using basis string}
In the protocol \ref{pro2}, we use four states to encode the evidence state without opening the basis of qubits. In fact, we can also encode the basis string of qubits while opening the string of qubits itself. Here we present a protocol follows this idea.

\begin{protocol}\label{pro5}~
\begin{enumerate}
\item Alice chooses a commitment value $b\in \{0,1\}$.
\item Alice chooses $a^{(i)}, b^{(i)}\in\{0,1\}^{n},i=1,2,\ldots,m$ randomly, $b^{(i)})$ satisfy $F(b^{(i)})=b$.
\item Alice prepares $m\times n$ qubits in state  $|a^{(1)}\rangle_{b^{(1)}}\cdots|a^{(m)}\rangle_{b^{(m)}}$ and sends them to Bob as a piece of evidence for her commitment.
\item Alice sends the values of $a^{(i)}$ within commit phase.
\item In open phase, Alice unveils the values of $b$ and $b^{(i)}$.
\item Bob checks each qubit via projective measurements as above.
\item Bob accepts the commitment value if $b^{(i)}$ satisfies $b=F(b^{(i)})$ for every $i$.$\hfill\Box$
\end{enumerate}
\end{protocol}

It is worth to mention that we can get $a_j^{(i)}$ from the qubit $|a_j^{(i)}\rangle_{b_j^{(i)}}$ and $b_j^{(i)}$, but we cannot get the value of $b_j^{(i)}$ from the qubit and $a_j^{(i)}$ with probability 1. Based on this property Alice can open $a^{(i)}$ before open phase, and the binding condition is still guaranteed with the aid of correlation immune Boolean function.

Note that if $a_j^{(i)}=0$ for every $i, j$, this protocol becomes the same as the protocol \ref{pro1}.

\subsection{Scheme using relative phase}
Besides using basis, we can also use a relative phase to commit.

\begin{protocol}\label{pro6}~
\begin{enumerate}
\item Alice chooses a commitment value $b\in \{0,1\}$, and chooses randomly $x, e\in\{0,1\}^{n}$ satisfying $e\neq(0,0,\cdots,0)$.
\item Alice prepares state $|x\rangle+(-1)^b|x\oplus e\rangle$ and sends the state to Bob.
\item Alice opens the values of $e$ and $b$.
\item Bob chooses randomly one nonzero bit of $e$, and uses the corresponding qubit as control qubit to do CNOT operation to qubits corresponding to other nonzero bits of $e$. After these Bob checks state of the control qubit by measuring it with basis $\{|+\rangle, |-\rangle\}$. He accepts the commitment, if the result is $b$.$\hfill\Box$
\end{enumerate}
\end{protocol}

Define \begin{eqnarray}
\rho_b^{(n)}=\frac{1}{2^n(2^n-1)}\sum_x\sum_{e\neq0}\left(|x\rangle+(-1)^b|x\oplus e\rangle\right)\left(\langle x|+(-1)^b\langle x\oplus e|\right).
\end{eqnarray}
It can be proved that $D(\rho_0^{(n)}, \rho_1^{(n)})<\frac{1}{p(n)}$. Therefore, Alice can prepare the following state if she wants to attack:
\begin{eqnarray}
\sum_{x,e}\frac{1}{2^{2n}}|x,e\rangle_A\otimes(|x\rangle+|x\oplus e\rangle)_B.
\end{eqnarray}

\subsection{An interactive scheme}
Let $F_1,F_2\cdots F_k$ be $k$ sets of Boolean functions, the domain of the function in $F_i$ is $\{0,1\}^{n_1+n_2+\cdots+n_i}$.
\begin{protocol}\label{pro7}~
\begin{enumerate}
\item Bob chooses randomly $f_{1j}\in F_1$ and sends it to Alice.
\item Alice chooses a commitment value $b\in \{0,1\}$ and chooses randomly $a^{(1)}, b^{(1)}\in\{0,1\}^{n_1}$ satisfying $f_{1j}(a^{(1)})=b$. Alice sends $|a^{(1)}\rangle_{b^{(1)}}$ to Bob.
\item Bob chooses randomly $f_{ij}\in F_i$ and sends it to Alice.
\item Alice chooses randomly $a^{(i)}, b^{(i)}\in\{0,1\}^{n_i}$ satisfy $f_{ij}(a^{(1)}, a^{(2)}, \cdots, a^{(i)})=b$, and sends $|a^{(i)}\rangle_{b^{(i)}}$ to Bob.
\item Repeat steps 3 and 4 with $i=2,\cdots,i_0$, here $i_0$ is chosen by Bob for each execution of the protocols.
\item Alice opens $b$ and all the states she has sent.
\item Bob checks the states.
\item Bob verifies that the output of every function he chose is $b$, and accepts the commitment.$\hfill\Box$
\end{enumerate}
\end{protocol}

In this protocol, if Alice wants to attack with the attack of the no-go theorem, she has to take into account all possible replies of Bob before the execution of the protocol, and prepares a state as follows:
\begin{eqnarray}
\sum_{j_1, b^{(1)}}\sum_{f_{1j_1}(a^{(1)})=0}\left(|a^{(1)}\rangle_{b^{(1)}}\otimes\sum_{j_2, b^{(2)}}\sum_{f_{2j_2}(a^{(1)}, a^{(2)})=0}\left(|a^{(2)}\rangle_{b^{(2)}}\otimes\cdots\cdots\right)\right).
\end{eqnarray}

In other protocols, Alice can prepare the state for each $i=1, 2, \cdots, m$ separately, but in this protocol, it is entangled for $i=1, 2, \cdots, m$. It seems more complex than that of other protocols, but can be prepared efficiently still.

\section{Practical scheme against channel loss and error}\label{sec8}
The protocols described above will be much useful if we can transform them into practical ones. Here we present a way to realize this goal by using error correcting code (ECC).\\

{\bf Channel loss}

One may think that the protocols proposed are already secure against channel loss, this opinion is based on such a consideration: Alice does not know which qubits are lost, then she cannot cheat via a different opening of these qubits successfully all the time. Bob can simply verify the consistency of his measurement results and the values Alice opened to decide whether to accept the commitment value.

In fact, a problem exists in every QS-based protocol executed over a lossy channel is that Alice can always attack with a low loss channel: she keeps several qubits in hand and sends the rest with a low loss channel, then she can cheat via opening these qubits with different values and Bob cannot detect this attack at all.\\

{\bf Channel error}

In this situation the QS-based protocol without additional design cannot be operated properly, since the inconsistency between the opened information and the measurement results can be owed to either channel errors or Alice's cheating.\\

{\bf The solution}

Generally speaking, channel loss can be regarded as a kind of channel error, because a disappeared qubit can always be regarded as an error qubit in state $|0\rangle$. Therefore, if a QS-based bit commitment protocol is one against channel error, we treat it as one against channel loss.

Next we construct a protocol based on ECC. In order to keep concealing, we should build the ECC $C$ as follows:

Suppose $\xi\times\eta$ matrix $G$ and $\eta\times(\eta-\xi)$ matrix $H$ are generator matrix and check matrix of an ECC $C_1$ with error correcting ability $t$, and there is one row of $H$ whose every entry is "1". It can be shown that any $2t$ rows of this matrix are linear independent.

Let $(\eta-1)\times(\eta-\xi)$ matrix $H'$ has every row of $H$ except that with all "1" entries. Define $C$ by a generator matrix $\widetilde{G}_{(\eta-\xi)\times(\eta-1)}=(H')^T$, here requiring $\eta-\xi$ is a factor of $n$. Then the check matrix of $C$ is $\widetilde{H}_{(\eta-1)\times(\xi-1)}$, and the $n$-qubit string is encoded into $\zeta$-qubit string, here $\zeta=\frac{n}{\eta-\xi}\times(\xi-1)$. Generally speaking, it is difficult to get the optimal check matrix from a generator matrix since this problem is related with the NP-complete problem of finding the decode algorithm of a general linear ECC. However, we can get the $\widetilde{H}$ efficiently for given parameters $\eta, \xi$.

The above method leads to that the values of any $2t-1$ bits of each codeword of $C$ are independent statistically from the commitment value. As a result, the probability of Bob's getting the parity bit with an $(\eta-1)$-qubit string is less than
\begin{eqnarray}
p_{max}^{(1)}=\sum_{i=2t}^{\eta-1}C_{\eta-1}^ip_s^i(1-p_s)^{\eta-1-i},
\end{eqnarray}
where $p_s$ denotes the probability of Bob getting one qubit's value correctly, which is related to the probability of distinguishing two nonorthogonal states and of channel error rate. Then the probability of Bob's getting the commitment value with an $\zeta$-qubit string is less than
\begin{eqnarray}
p_{max}=(p_{max}^{(1)})^{\frac{n}{\eta-\xi}}.
\end{eqnarray}

As the number of $\zeta$-qubit-strings involved in a protocol is $m$, the probability of Bob's getting the commitment value is less than
\begin{eqnarray}
P_{max}=1-(1-p_{max})^m.
\end{eqnarray}

Assume the probability of channel error is $p_{ce}$, for any QS-based protocol with $n_{th}$ evidences, the worst situation is that Alice has a super channel with no channel error and then she can open with some values changed which are chosen by her. If the changes Bob found are less than $n\times p_{ce}$, Alice can cheat successfully. However we show that the encoding with suitable ECC $C$ can help Bob resist Alice's attack and benefits the binding condition.

Let the error correcting ability of $C$ is $t'$, it satisfies
\begin{eqnarray}
t'>(\xi-1)\times p_{ce}.
\end{eqnarray}
Assume each change of value by Alice should be found out by Bob with probability $p_{cv}$(it should be $\frac{1}{2}$ in most cases), then we just need
\begin{eqnarray}
(t'+1)\times \frac{n}{\eta-\xi}\times p_{cv}>n\times p_{ce}.
\end{eqnarray}
This means if the expanded protocol with ECC satisfies
\begin{eqnarray}
t'>(\eta-\xi)\times\frac{p_{ce}}{p_{cv}}-1,
\end{eqnarray}
every time Alice cheat with value changes, the number of error Bob found will be more than it should be. Therefore the protocol can resist the super channel attack by Alice.

However, this method leads to redundant information, which is disadvantageous to the concealing condition. We need the protocol satisfies the Theorem \ref{the1}.

Take the protocol \ref{pro2} for an example. Assume $m=1$, then the extended protocol is shown as follows:
\begin{protocol}\label{pro8}~
\begin{enumerate}
\item Alice makes a commitment $b\in \{0,1\}$.
\item Alice chooses $a=(a_1,a_2\ldots,a_n)\in\{0,1\}^{n}$ randomly. Then she uses ECC $C$ to code $a$ and gets $c^1=(c_1^1, \cdots, c_{\xi-1}^1),\cdots,c^{\frac{n}{\eta-\xi}}=(c_1^{\frac{n}{\eta-\xi}}, \cdots, c_{\xi-1}^{\frac{n}{\eta-\xi}})$.
\item Alice chooses $b_1^1,\cdots,b_{\xi-1}^{\frac{n}{\eta-\xi}}\in\{0,1\}$ randomly, prepares $\zeta$ qubits in state $|c_1^1\rangle_{b_1^1}\cdots|c_{\xi-1}^{\frac{n}{\eta-\xi}}\rangle_{b_{\xi-1}^{\frac{n}{\eta-\xi}}}$
    and sends to Bob as a piece of evidence for her commitment.
\item Alice opens by declaring $b$ and the values of $a$, $c_i^j$ and $b_i^j$.
\item Bob checks states of qubits by corresponding projective measurements: if $b_i^j=0$, Bob measures with basis $\{|0\rangle, |1\rangle\}$, else with basis $\{|+\rangle, |-\rangle\}$. Bob decode the result of measurement with $C$, the error probability should less than $p_{ce}$ and the message should be $a$.
\item Bob checks value $b=F(a)$.  \hfill$\Box$
\end{enumerate}
\end{protocol}

Assume $\varsigma_b$ is the density operator of quantum state Bob receives before open phase while Alice commits $b$, it should contain the channel error. Let the decoding process of $C$ be $C'$, so we get
\begin{eqnarray}
\varsigma_b=\sum_{F(C'(c^1),\cdots,C'(c^{\frac{n}{\eta-\xi}}))=b}\left(\bigotimes_{j=1}^{\frac{n}{\eta-\xi}}\bigotimes_{i=1}^{\xi-1}|c_i^j\rangle_{b_i^j}\langle c_i^j|\right)
\end{eqnarray}
they should satisfy that
\begin{eqnarray}
D(\varsigma_0,\varsigma_1)<\frac{1}{p(n)}.
\end{eqnarray}

\section{Discussion}

Our analysis of security above is for the situation that only one of Alice and Bob is dishonest.

It can be seen that the QS-based QBC protocols have a common weakness: the security of binding would not be guaranteed in a practical case with channel loss or error. We solve this problem for the first time via transforming QS-based protocols into ones with error correction coding. Both bounded channel loss and error can be solved in this way, since we can take channel loss as a special channel error and operate against it with error-correction-code(ECC). Only the conditions for the QBC protocols based on parity function have been given explicitly. How to transform general QBC protocols into practical ones is still worth considering. Furthermore, the ECC-based method is proved secure against individual attack only, the security against more general attacks is still an open problem.

Another problem in practice is the lack of single photon source. It can be seen that the weak coherent pulse source cannot guarantee the two necessary conditions at the same time: 1. Alice sends almost every qubit via emitting single photon; 2. Bob receives almost every qubit. It can be seen that the weak coherent pulse source is not suitable for our protocol. We need a single photon source to accomplish the practical protocol in some laboratories.

\section{Conclusion}
We suggest a definition of information-theoretical concealing for quantum bit commitment, then propose three kinds of QS-based bit commitment protocols and prove that they are information-theoretically concealing. The binding of them is considered under a new concept "physical security of protocol".

We have also suggested other four QS-based protocols without proof of security. They will give some hints to help us get closer to the goal of unconditionally secure QBC protocol.

Finally, we give a method to transform QS-based protocols into practical ones with ECC.

\section*{Acknowledgment}
This work was supported by National Natural Science Foundation of China under Grant No. 60573051.





\bibliographystyle{model1a-num-names}
\bibliography{<your-bib-database>}






\appendix

\section{The attack to the binding condition}\label{appa}
From Eq.(\ref{eq2}) we have
\begin{eqnarray}
F(\rho_0^{(n)},\rho _1^{(n)})\geq1-(\sin\alpha)^n.
\end{eqnarray}
As shown in Sec.\ref{subsec1}, this result means Alice can use a local unitary transformation to perform a successful cheat. Here we show a method to solve out $U_A$ of QS-based protocols.

For a cheating Alice, the states she prepared were shown as Eqs.(\ref{eqhat0}) and (\ref{eqhat1}) in Sec.\ref{subsec1}. Now Alice needs to get the state $|\nu\rangle$ whose reduced density operator is the same as that of $|\hat{0}\rangle$, and satisfies $\langle\hat{1}|\nu\rangle=F(\rho_0^{(n)},\rho _1^{(n)})$. After that she must find out the local unitary transformation $U_A$ to transform $|\hat{0}\rangle$ into $|\nu\rangle$.

In order to achieve these goals, Alice should do as follows:
\begin{enumerate}
\item The Schmidt decomposition of $|\hat{0}\rangle$ and $|\hat{1}\rangle$.

There exists an orthogonal basis set $\{|0\rangle, |1\rangle\}^{\otimes n}$ for subsystems A and B, thus $|\hat{0}\rangle$ can be written as
\begin{eqnarray}
|\hat{0}\rangle=\sum_{i,j}\theta_{ij}|i\rangle_A\otimes|j\rangle_B ,
\end{eqnarray}
where $i,j\in\{0,1,\ldots,2^n-1\}$, and
\begin{eqnarray}
\theta_{ij}=\sum_k\alpha_k\ {_B}\langle j|\varphi_j^{(0)}\rangle_B\nonumber
\end{eqnarray}
if $w(i),w(j)$ is even, here $w(\cdot)$ means Hamming weight; else $$\theta_{ij}=0.$$

Let $\Theta$ be a matrix with entries $\theta_{ij}$. According to the singular value decomposition, we have $\Theta=UDV$, here $D$ is a diagonal matrix with positive elements, and $U$ and $V$ are unitary matrices.
Thus
\begin{eqnarray}
|\hat{0}\rangle=\sum_{i,j,k}u_{ik}d_{kk}v_{kj}|i\rangle_A\otimes|j\rangle_B.
\end{eqnarray}
Define $|x_k\rangle_A=\sum_i u_{ik}|i\rangle_A$, $|y_k\rangle_B=\sum_j v_{kj}|j\rangle_B$, and $\lambda_k=d_{kk}$, we can see that
\begin{eqnarray}
|\hat{0}\rangle=\sum_k\lambda_k|x_k\rangle_A\otimes|y_k\rangle_B.
\end{eqnarray}
It can be seen that $\{|x_k\rangle_A$\}, $\{|y_k\rangle_B\}$ form two orthogonal basis sets.

Similarly, Alice gets
\begin{eqnarray}
|\hat{1}\rangle=\sum_k{\lambda'}_k|{x'}_k\rangle_A\otimes|{y'}_k\rangle_B.
\end{eqnarray}
\item The polar decomposition of $\sqrt{\rho_1^B}\sqrt{\rho_0^B}$.

$\rho_0^B$ and $\rho_1^B$ are defined with Eq. (\ref{eq3}), the related polar decomposition is
\begin{eqnarray}
\sqrt{\rho_1^B}\sqrt{\rho_0^B}=\Big|\sqrt{\rho_1^B}\sqrt{\rho_0^B}\Big|T.
\end{eqnarray}
There exists an orthogonal basis set with which $\rho_0^B$ and $\rho_1^B$ are in block-diagonal form\cite{Ben96} and the blocks have a general expression, so that we can give the entries of matrix $T$ based on this orthogonal basis.
\item Solving $U_A$.

Based on the proof of Uhlmann's theorem given by Jozsa \cite{Joz94, Nie02}, we have
\begin{eqnarray}
|\nu\rangle=\Big(I\otimes\sqrt{\rho_0^B}T^{\dag}\Big)\sum_i|x'_i\rangle_A\otimes|y'_i\rangle_B.
\end{eqnarray}
It can be seen that there exists a local unitary transformation of Alice, or $U_A$, transforming $|\hat{0}\rangle$ into $|\nu\rangle$.

Note that $\rho_0^B=\sum_i |\lambda_i|^2|y_i\rangle_B{_B}\langle y_i|$, it gives
\begin{eqnarray}
|\nu\rangle&=&\left(I\otimes\sqrt{\rho_0^B}T^{\dag}\right)\sum_i|x'_i\rangle_A\otimes|y'_i\rangle_B \nonumber\\
&=&\sum_{i,j}|x'_i\rangle_A\otimes\lambda_j|y_j\rangle_B{_B}\langle y_j|T^{\dag}|y'_i\rangle_B\nonumber\\
&=&\sum_j\lambda_j\left(\sum_i{_B}\langle y_j|T^{\dag}|y'_i\rangle_B|x'_i\rangle_A\right)\otimes|y_j\rangle_B.
\end{eqnarray}

It can be seen that
\begin{eqnarray}
U_A|x_i\rangle=\sum_i{_B}\langle y_j|T^{\dag}|y'_i\rangle_B|x'_i\rangle_A .
\end{eqnarray}
Then Alice can get all elements of $U_A$ from this equation.
\end{enumerate}

\end{document}